\documentclass[a4paper]{article}

\usepackage{graphicx}
\usepackage{hyperref}

\author{Ana Barroso - abarroso@seemoo.tu-darmstadt.de}
\title{aDTN - Undetectable Communication \\ in Wireless Delay-tolerant Networks\\ \vspace{0.5cm} \Large Working Draft\footnote{The most recent version of this document can be found at \url{http://www.seemoo.de/team/ana-barroso/adtn/}}}

\begin{document}

\maketitle

This document describes a best-effort delay-tolerant communication system that protects the privacy of users in wireless ad-hoc networks by making their communication undetectable. Delay-tolerant communication takes place when end-to-end connections between any two nodes of the network may not exist. This means that the network might split at several places and at different points in time, perhaps with nodes travelling between network partitions to pass messages along in a store-carry-and-forward approach.

The proposed system is a wireless broadcast-based adaptation of mix networks~\cite[p.~6]{anonymity-survey} where each user belongs to at least one group it trusts, and each group acts as a mix node\footnote{A mix node is an intermediate node that collects and shuffles messages before sending them to the respective recipients. The goal of mixed networks is to hide the link between the sender and the recipient of a message by routing them through mix nodes.}. Assuming encryption is not broken, it provides undetectability of all users and messages against external adversaries, as well as undetectability of users and messages in non-compromised groups against internal adversaries.

The document is structured as follows: we start by giving a short introduction to privacy terminology in Section~\ref{term}. In Section~\ref{goals} we describe the privacy protection goals of the system. In Section~\ref{system} we describe the system, by listing and explaining its requirements. In Section~\ref{threat} we analyse the provided privacy protection against various attacker types. New comments to the system and proposed extensions are listed in Section~\ref{comments}. Finally, in Section~\ref{futurework}, we describe the next steps for this work.

%\tableofcontents

\section{Privacy terminology}
\label{term}
We use the privacy terminology by Pfitzmann and Hansen~\cite{anonymity-term}; in this section we summarize the relevant terms for this work:

\textsl{Unlinkability} is the probability of a particular attacker not being able to sufficiently distinguish a link between two items of interest within a set, the \textsl{anonymity set}. All the following terms are a specific form of unlinkability:

\textsl{Sender anonymity} is the unlinkability between a received message and its original sender. \textsl{Recipient anonymity}, analogously, is the unlinkability between a sent message and its intended recipient.

\textsl{Relationship anonymity} is the unlinkability between the original sender of a message and the intended recipient.

\textsl{Undetectability} provides relationship anonymity while also hiding the act of sending and the act of receiving. The sender is not linked to the messages it sends, and the recipient is not linked to the messages it receives. This allows both sender and receiver the possibility of invoking plausible deniability, as no attacker can certainly be sure that the subject being observed is taking part in the communication at all.

\textsl{Unobservability} extends undetectability by also requiring the anonymity of the item of interest (which could be a user or a message) against every entity that interacts with it, and not only against attackers.

\section{Privacy protection goals}
\label{goals}
Our system has the goal of ensuring that, in a wireless delay-tolerant network, \textsl{communication is undetectable} by untrusted entities: an entity not trusted by a node cannot estimate whether a message sent or received by that node has content or is completely random, as the two cases appear to be equally likely.

We note that such a system with S sending nodes and R receiving nodes will also ensure the following properties:
\begin{itemize}
\item that for every message transmitted in the network, the corresponding sender anonymity is $1-\frac{1}{S}$ against entities the sender does not trust;
\item that for every message transmitted in the network, the corresponding recipient anonymity is $1-\frac{1}{R}$ against entities the recipient does not trust;
\item  and therefore that the real sender and the intended recipient of a message are completely indistinguishable from all other potential senders and recipients. The system provides the highest unlinkability between sender and recipient of a message.
\end{itemize}

\section{System description}
\label{system}

The system described in this section is an adaptation of mix networks to a wireless ad-hoc scenario. Mix networks can at most guarantee sender and recipient anonymity, and therefore relationship anonymity, but they do not guarantee unobservability, nor do they guarantee undetectability against attackers at the endpoints of the mix network. Apart from that, they have been designed for wired networks, where eavesdropping by an attacker takes a completely different approach.

We do not describe here all the issues that a standard wired mix would have in a mobile wireless network, but instead focus on the approach we have devised, where all participant nodes behave similarly to a mix node.

Note that what we describe here is a network layer protocol that hides the act of sending or receiving messages. Messages that come from the above layers may or may not be encrypted. If they are not encrypted, their contents may be read by every user in the network; this can be useful to broadcast public information. If users wish to send a message while ensuring sender anonymity, they must take care not to include any information in that message that could possibly identify them.

In order to achieve the goal of undetectable communication, we established the following requirements:
\begin{description}
\item[Untraceable] \textsl{A message has to look different at each hop.}
\\ An entity that is not trusted by a given group of nodes will only be able to see that they transmit completely indistinguishable, random messages of equal size, and this at constant intervals %TODO
in time. The reason why a message has to look different at each hop is to prevent an attacker from tracing the path of the message back to the original sender. If an attacker has the capability to collect and analyse the whole traffic since the network started, it can determine who was the first user to ever transmit the message. However, when all messages look different between hops, anyone could be the message originator, thus the original sender's anonymity set is the whole network.

\item[Indistinguishable] \textsl{All messages have the same size and do not carry any address or other routing information.}
\\ All transmitted messages must be indistinguishable from each other. Therefore they cannot contain any information that could help an attacker to identify it and track it back to its source. For example, they cannot contain addresses, nor time-to-live or hop count fields. Analysing message sizes across hops can also lead to their correlation.

\item[Trust group] \textsl{A group of users that trust each other exchange a symmetric key. They can do it beforehand or at the place where they meet, using another (secure) medium.}
\\ In this scheme, a mix is a group of nodes, and not just one node. The idea here is that only users who trust each other share a key. If one of them wants to include another member in the group that is not trusted by all participants, they should start a new group. As a mix node \textsl{trusts itself}, in the same sense a distributed ``group-mix'' node trusts itself. A compromised mix node will not provide as much privacy as a healthy one, and the same applies to our distributed ``group-mix". The damage is limited, however: an attacker will only be able to determine the direct contacts of that node, and can only track messages within the groups that the node belongs to.

\item[Multiple groups] \textsl{If a particular user trusts more than one group, she can have multiple keys, one for each group.}
\\ When a node is in different groups, its acts as a bridge between mixes. In the original mix scenario, a mix node would encrypt and address a message to another mix node. In this case, it is as if two mix nodes overlap in a single user node: once a node belonging to multiple groups possesses a message, it can distribute the message through all the groups it belongs to.

\item[Sending messages] \textsl{When a node needs to send a message, it computes the message's fingerprint and attaches it to the message. For each owned key, it encrypts the message before adding it to the pool of outgoing messages.}
\\ Since messages do not carry a group identifier, the nodes must have a way to find out if an attempted decryption process was successful or not. We propose to attach a fingerprint to the message before encrypting it in order to achieve this. How the fingerprint is computed depends on the implementation. The sending process is depicted in Figure~\ref{nodes_layers}.

\begin{figure}
\includegraphics[width=\textwidth]{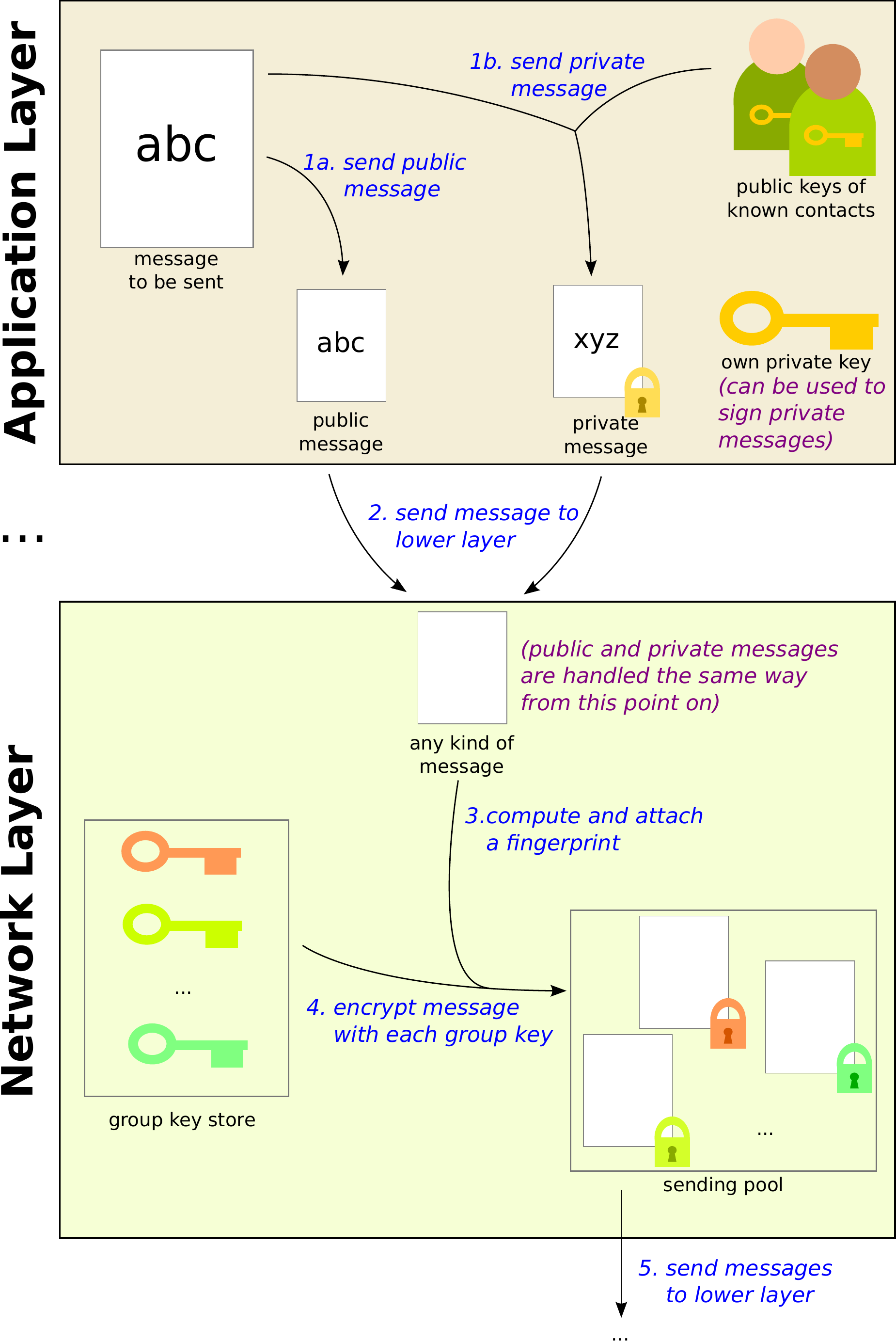}
\caption{A user writes a message and leaves it in plaintext if it is to be sent to all members of the network, or encrypts it for a single user. The message is passed to the network layer, where the anonymisation process takes place: first a fingerprint, such as a checksum, is attached, and for each group key, a copy of the message is encrypted and added to the pool of outgoing messages, to be later transmitted.}
\label{nodes_layers}
\end{figure}

\item[Receiving messages] \textsl{When a node receives a message, it tries to decrypt it with each of the owned keys, until one of them is successful and the message becomes readable, or until all fail.}
\\ Unless the node has a way to determine if the transmitter of that particular message is in a trusted group, the node will have to verify the decrypted message with the bits that contain the fingerprint.

\item[Discarding unreadable messages] \textsl{If decrypting a received message is not successful with any key, the message is discarded.}
\\ A message is useless for a node that cannot decrypt it. It cannot be reencrypted, as a node that could potentially decrypt it would have to go through at least two layers of decryption and might not have both keys, nor has any indication that there is more than one layer.

\item[Ensuring message freshness] \textsl{If the message is readable (because decryption was successful), the node determines if the message is old. If so, it is discarded. Otherwise, the node passes it to the network layer above and also adds the message to the sending pool to be forwarded further.}
\\ Once a node determines that the message is readable, it might be the case that the message is too old to be retransmitted, so the node does not need to retransmit it further. Since the message does not carry any time-to-live counter nor any sort of timestamp, the node itself must decide when a message is too old. This can be done by keeping a log of the first time the message was seen and how many times it was received from other nodes, for example.

\item[Forwarding messages] \textsl{If a message is to be forwarded, it is sent as if it was a new message (see requirement ``Sending messages'').}
\\ Every message is sent the same way, whether it is an original message or a retransmitted one. This way, no one who hears the transmission, not even another member of the same mix, can know if that user is the originator of the message or not.

\item[Cover traffic] \textsl{Messages are sent from the pool of outgoing messages with a constant frequency. If there are no messages to be sent, a random message, indistinguishable from all other messages, is transmitted.}
\\ According to ~\cite[p.~8-10]{anonymity-survey}, classical mixes can be vulnerable to flooding and timing attacks, whose goal is to link a message received by the mix to the corresponding message at the next hop. Several buffer flushing strategies have been introduced to counter such attacks. However, this kind of attacks have no impact here, since our nodes discard messages from groups they do not belong to (and if the attacker is in the same group, it already knows the information it could get by performing the attacks). We define transmission frequency to be constant so that all nodes have indistinguishable behaviour, regardless of how many outgoing messages they have in their buffer.

\end{description}

From these characteristics it follows that the size of the groups should be kept small, and a node should belong to as few groups as possible, in order to minimise the damage of a compromised node. However, this can negatively impact the exchange of messages between groups and therefore the overall traffic flow. We will analyse the impact of such parameters at a later stage.

Regarding the previously mentioned protection goals (Section~\ref{goals}), we can consider $S=R=N$ (number of nodes in the network), as all nodes are constantly sending traffic and are at the same time the potential recipient of any message.

%\subsection{Cover traffic}

%According to~\cite{counter-stat-disclosure}, in a statistical disclosure attack, ``the attacker isolates his attack against a single user, which we will call Alice. The statistics used in this attack are the frequencies with which each recipient gets a message from the system. By taking differences between the frequencies observed when Alice is active and those observed when she is not active, the attacker can estimate Alice’s contribution to the recipient set.'' In our system, it is not possible the observe when a node receives a message, only when it sends.

\section{Analysis of privacy protection effectiveness}
\label{threat}
In this section we analyse how well the system fares against different kinds of adversaries. We start by analysing external adversaries and then internal ones. We not only analyse privacy effectiveness, but also the censorship resilience of the network. We conclude the section with a summary of the achieved privacy protection.

\subsection{External adversaries}
External adversaries do not play a role in the communication system being analysed. In this case, that means they do not control any of the nodes in the network.

If the adversaries are only \textbf{passive}, they can monitor and store the traffic being exchanged. Since all the messages are indistinguishable from random noise, they cannot correlate traffic to users, which then have perfect anonymity. However, they can attempt to decrypt it at a later stage. The time they need for this mostly depends on their computing power and the encryption strength. If the adversaries can decrypt all the traffic in the whole network (\textbf{global} eavesdropper), they might be able to discover all the relationships between users and messages. \textbf{Local} adversaries are limited in this, unless they are also \textbf{mobile}, but even then they will only have a few snapshots of the communication over time. A plausible scenario for this would be to specifically monitor a particular user.

If the adversaries are \textbf{active}, they can emit radio signals. If the users try to interpret the signals as legitimate messages, they will discard them as unreadable, as it is highly unlikely that one of their keys can turn the received frame into a message and the corresponding fingerprint. In this case, if the emissions are sporadic, there is little damage done. However the attackers can perform denial-of-service (DoS) attacks, since they can jam a node or a group of nodes if they constantly transmit a signal, and also cause them to drain their batteries in pointless decryptions. Again, if the adversaries are \textbf{mobile}, they can target particular subjects. A sufficiently powerful global adversary could jam the whole network, effectively preventing all the nodes from communicating. While DoS is not a direct attack to privacy, it might lead the users to resort to less secure channels of communication.

\subsection{Internal adversaries}

Internal adversaries can perform all the attacks of external adversaries, and in addition they can attempt to obtain more information from their position in the network. In our scenario, an internal adversary can control one or more nodes possessing group keys shared with non-adversarial users, either as their legitimate user or by capturing the devices of other users. It would be useless for the attackers to only share keys among themselves and not with legitimate users: in that case they would be external adversaries, as they would not be connected to the target network.

As all messages look different at each hop, it makes no difference to be an \textbf{active} or \textbf{passive} attacker in terms of tracking a message along its path. The only advantage of being an active attacker is to be able perform a DoS attack by flooding the network with unreadable messages, but in this case it behaves as in the scenario of the external jammer, which we already covered.

A \textbf{local} attacker can only directly communicate with or monitor the nodes that also own the keys she has in her possession. If the attacker personally exchanged the keys with the other users, she does not gain any information, as she already knew who she is communicating with. However, if she captured the node from another user, she may find out who trusts whom (because they shared a group key), thus revealing parts of the users' social graph. While she can see who is rebroadcasting which messages, she cannot prove that they are the original senders or the intended recipients of the messages.

If several internal attackers collude, they can expose a larger portion of the social graph. In an extreme scenario, where all groups are infiltrated by an adversary who has been monitoring the network since the beginning of communication (in which case we can say the adversay has a \textbf{global} view of the network), it can determine the original senders of a given message. While the scenario of having spies in each group seems quite unlikely, it could be possible for an adversary to capture all the devices in the network (e.g. authorities seize the mobile phones of all citizens in a protest).

\subsection{Summary of privacy analysis}

The system achieves the goal of communication undetectability we described in Section~\ref{goals} and goes even further in ensuring privacy protection: When a message is transmitted, only the users within the group sharing the corresponding key will know which of the members of the group first transmitted it \textsl{within that same group}, but they cannot be sure that user is the message originator: she could have received the message through another group of users. Moreover, not even the members in the trust group of the intended recipient will know that the message is meant for her, meaning that receiving a message is unobservable.

If we consider a group to be an entity, we can argue that the system also ensures \textbf{unobservability} of that group: all the users in the group and all the messages they transmit among each other are undetectable \textsl{against the users that do not belong to the group}, even if they are involved in the transmission of the same message.

Assuming that adversaries cannot break the encryption of the messages, the system ensures that the act of sending a message is \textbf{undetectable} \textsl{against all kinds of external adversaries}, and also \textsl{against internal adversaries} as long as the message is being sent in a group that has not been compromised. However, the internal adversaries cannot prove that a user is the originator of a message, unless they have compromised every existing group.

If the adversaries have enough computing power to break the encryption of all overheard messages and have a global view of the network, they may be able to reconstruct the social graph of all network users and determine the original sender - but not the intended recipient - of every message; if they can only decrypt a few messages or have partial view of the network they can at most obtain a limited view of the said graph and cannot determine the source of a message. We point out that \textbf{protecting the cryptographic material from being read by third parties is out of the scope} of our system. How effectively it is done depends on other mechanisms which we do not discuss here, such as operating system security.

\section{Comments and proposed extensions}
In this section we list comments and new ideas we received since the last version of the document. Please feel free to discuss them on the mailing list\footnote{\url{adtn-devel@lists.sourceforge.net}}, you're most welcome to do so..
\label{comments}
\begin{description}
\item[Extend coverage with a mesh network] A cooperating mesh network could help improve the delivery ratio and time. A mesh node receiving aDTN traffic could broadcast it through the mesh network. The other nodes would rebroadcast the aDTN messages. Since the mesh nodes are not able to decrypt the traffic, it would have to be retransmitted unchanged. This makes the mesh network susceptible to DoS attacks; a solution could be to limit the aDTN traffic to a certain bandwidth. %idea by me

\item[Preventing rogue nodes from joining a group] According to the scheme, nodes that trust each other share a symmetric key to form a mix group. However, it is possible that a user betrays the group and shares the key with other users. We wonder if there is way to prevent this, e.g. by making it mandatory that all group members trust the user attempting to obtain the key. Such a system ``enforces anonymity instead of relying on the users behaving flawlessly, or never being pressured into adding users untrusted by most of the group (think coercion)". If the reader has an idea on how to do this, please let us know. % idea by Andrés Valoud

\item[Whitelisting and blacklisting] Right now we have a very inefficient way of checking if we can decrypt a received message or not. One possibility is to store the mac addresses of the nodes that sent messages we were able to decrypt, but most of the received messages are probably from nodes of other groups. It was suggested to also make a blacklist with the nodes that sent at least T messages we were unable to decryp, where T is some user-defined or system-defined threshold. To prevent the list fromg growing too large, maybe we can also store when was the last time we heard each node and remove it from the list after some time. % idea by Andrés Valoud

\item[Magic number for early identification] Another suggestion to check if a message is going to be decrypted is ``to allow the system to detect and abort the decryption of a negative at a much earlier stage, and at a predictable speed. It works by adding a magic number (1-4 bytes) to the beginning of a packet before it is encrypted and sent to the other nodes. This will allow the receiving nodes to identify and abort full decryption of a negative at a much earlier stage than the current solution''. Since this may make it easier for an attacker to break the encryption, we will have to discuss it thoroughly and decide if it is really necessary. % idea by Mikael Turøy

\item[Countering spam] There is an insider attack we missed: what if nodes in the network start spamming many messages by transmitting more often than other nodes? How can we prevent them from flooding the network? Should listening nodes also register the transmission rates of their neighbours? What if the attacker is highly mobile and covers the network really fast? Well, one option is to let the ``age checker" deal with it: instead of trashing a message after a certain number of retransmissions, we could add another filter, based on how often a message was overheard. If everyone is retransmitting the spam, it should not survive for long in the network. Another suggestion is to split the network to isolate the spammers, but the problem is how to do this: a node in a group with a spammer would have to leave that group, limiting its communication possibilities. %It was also suggested to intoduce a third type of message, the "no-op-message (something like ping, to gain spam resistance)", as, "paradoxally, increasing valid traffic makes it much harder for a spammer to succeed".  % idea by Andrés Valoud

\item[Unclogging the network] We received a remark that ``the sending frequency determines the available bandwidth in the network [...] how can the network change the sending frequency to accomodate varying actual traffic? That is, if the network becomes clogged, how does it unclog? And how is the act of unclogging made invisible?". A possible solution is to always send the less popular message stored on the device, i.e. the one we transmitted and/or overheard the least times. This is something we need to test in a simulation. % idea by Andrés Valoud

\end{description}

\section{Next steps}
\label{futurework}
Being a working draft, this document will be updated as the system design progresses. The goal is to write the technical specifications for a smartphone application and then implement it.

In this document we specify that nodes should transmit at a constant rate. However, we should make sure that this is the best way to construct cover traffic (another option could be to transmit at a random rate averaging a system-defined value).

The system can be customised with several parameters, such as group size, maximum number of groups a user can belong to, message freshness (how long a node will retransmit a message since it was first seen), transmission frequency and maximum message size. The values of these parameters may greatly impact network performance, which we can analyse with metrics such as throughput, time to delivery (in the case of a single intended destination), network diffusion rate (time needed to make a message reach a certain proportion of the network), etc. It is important to understand this impact, so our next step is to run simulations of the system on networks of different sizes and mobility rate and measure the network performance.

Additionally, at first sight the system seems to be very inefficient: every time a message is received, several decryption attempts must be performed, and at most one will be useful; nodes can be constantly receiving and transmitting messages, many of them redundant. These two aspects can considerably shorten node uptime, as batteries will drain faster than when the mobile devices are in idle mode. The effect of cryptographic operations and transmission frequency on battery life will also have to be analysed, and we will attempt to optimise the system further, without weakening the privacy protection.

\bibliographystyle{plain}
\bibliography{bibliography}

\end{document}